\documentclass[12pt,a4paper]{article}

\usepackage{amsmath,amssymb}
\usepackage[nosort]{cite}
\usepackage[hyperref,parsep]{collect}

\makeatletter
\let\old@makecaption=\@makecaption
\def\@makecaption{\small\old@makecaption}
\makeatother

\makeatletter
\let\old@startsection=\@startsection
\renewcommand{\@startsection}[6]{\old@startsection{#1}{#2}{#3}{#4}{#5}{#6\mathversion{bold}}}
\makeatother

\let\oldPhi=\Phi
\let\oldPsi=\Psi
\let\oldGamma=\Gamma
\let\oldDelta=\Delta
\let\oldSigma=\Sigma
\let\oldTheta=\Theta
\let\oldPi=\Pi
\renewcommand{\Phi}{\mathnormal{\oldPhi}}
\renewcommand{\Psi}{\mathnormal{\oldPsi}}
\renewcommand{\Gamma}{\mathnormal{\oldGamma}}
\renewcommand{\Sigma}{\mathnormal{\oldSigma}}
\renewcommand{\Delta}{\mathnormal{\oldDelta}}
\renewcommand{\Theta}{\mathnormal{\oldTheta}}
\renewcommand{\Pi}{\mathnormal{\oldPi}}

\newcommand{\smat}{\mathcal{S}}

\newcommand{\superN}{\mathcal{N}}

\newcommand{\Li}{\mathop{\mathrm{Li}}\nolimits}

\newcommand{\order}[1]{\mathcal{O}(#1)}

\newcommand{\Integers}{\mathbb{Z}}
\newcommand{\Reals}{\mathbb{R}}

\ifx\genfrac\sdflkaj

\else

\fi

\newcommand{\indup}[1]{_{\mathrm{#1}}}

\newcommand{\alg}[1]{\mathfrak{#1}}

\newcommand{\lrbrk}[1]{\left(#1\right)}
\newcommand{\bigbrk}[1]{\bigl(#1\bigr)}

\newcommand{\xexp}[2]{x^{#1}_{#2}}
\newcommand{\xp}[1]{\xexp{+}{#1}}
\newcommand{\xm}[1]{\xexp{-}{#1}}
\newcommand{\xpm}[1]{\xexp{\pm}{#1}}

\makeatletter
\newcommand{\ellSN}{\mathop{\operator@font sn}\nolimits}
\newcommand{\ellCN}{\mathop{\operator@font cn}\nolimits}
\newcommand{\ellDN}{\mathop{\operator@font dn}\nolimits}
\newcommand{\ellAM}{\mathop{\operator@font am}\nolimits}
\newcommand{\ellK}{\mathop{\smash{\operator@font K}\vphantom{a}}\nolimits}
\newcommand{\ellE}{\mathop{\smash{\operator@font E}\vphantom{a}}\nolimits}
\makeatother

\newcommand{\nln}{\nonumber\\}

\newcommand{\earel}[1]{\mathrel{}&\hspace{-2\arraycolsep}#1\hspace{-2\arraycolsep}&\mathrel{}}
\newcommand{\eq}{\earel{=}}

\def\[{\begin{equation}}
\def\]{\end{equation}}
\def\<{\begin{eqnarray}}
\def\>{\end{eqnarray}}

\newcounter{enumlistcnt}
\renewcommand{\theenumlistcnt}{\protect\emph{\roman{enumlistcnt}}}

\makeatletter
\def\mr@ignsp#1 {\ifx\:#1\@empty\else #1\expandafter\mr@ignsp\fi}%
\newcommand{\multiref}[1]{\begingroup
\xdef\mr@no@sparg{\expandafter\mr@ignsp#1 \: }%
\def\mr@comma{}%
\@for\mr@refs:=\mr@no@sparg\do{\mr@comma\def\mr@comma{,}\ref{\mr@refs}}%
\endgroup}
\makeatother

\newcommand{\hypref}[2]{\ifx\href\asklfhas #2\else\href{#1}{#2}\fi}

\newcommand{\appref}[1]{App.~\multiref{#1}}

\renewcommand{\eqref}[1]{(\multiref{#1})}

\ifx\href\asklfhas\newcommand{\href}[2]{#2}\fi
\newcommand{\arxivno}[1]{\href{http://arxiv.org/abs/#1}{#1}}

\begin{document}

\begin{flushright}\footnotesize
\texttt{\arxivno{hep-th/0606214}}\\
\texttt{PUTP-2200}
\end{flushright}
\vspace{0cm}

\begin{center}
{\Large\textbf{\mathversion{bold}%
On the Scattering Phase for $AdS_5\times S^5$ Strings%
}\par}
\vspace{1cm}

\textsc{Niklas Beisert}
\vspace{5mm}

\textit{Joseph Henry Laboratories\\
Princeton University\\
Princeton, NJ 08544, USA}\vspace{3mm}
\vspace{3mm}

\texttt{nbeisert@princeton.edu}\par\vspace{1cm}

\textbf{Abstract}\vspace{7mm}

\begin{minipage}{12.7cm}
We propose a phase factor of the worldsheet S-matrix 
for strings on $AdS_5\times S^5$
apparently solving Janik's crossing relation exactly.
\end{minipage}

\end{center}
\vspace{1cm}
\hrule height 0.75pt
\vspace{1cm}

\section{Introduction}

The discovery of integrability 
in planar AdS/CFT \cite{Minahan:2002ve,Beisert:2003tq,Bena:2003wd}
has given hope that both par\-tici\-pa\-ting models,
$\superN=4$ gauge theory and string theory on $AdS_5\times S^5$,
can be solved exactly in the planar limit.
The spectrum can be obtained, 
at least to the leading few orders in perturbation theory,
by asymptotic Bethe ans\"atze, see \cite{Beisert:2004ry} for a review.
The leading weak-coupling order in gauge theory was solved in \cite{Minahan:2002ve,Beisert:2003yb}.
Reliable Bethe equations presently exist for up to second order (three loops)
\cite{Serban:2004jf,Beisert:2004hm,Beisert:2005fw}.
The spectrum of classical string theory was solved in 
\cite{Kazakov:2004qf,Beisert:2005bm} by means of spectral curves.
Based on these results, Bethe equations for quantum strings were
proposed \cite{Arutyunov:2004vx,Beisert:2005fw}.
The current state of the art is the expansion to first strong-coupling order 
\cite{Beisert:2005cw,Schafer-Nameki:2005is,Hernandez:2006tk,Freyhult:2006vr}. 

The Bethe equations for gauge theory can be derived by means of
an asymptotic Bethe ansatz \cite{Staudacher:2004tk}. 
This ansatz transforms the spin chain states 
into a one-dimensional particle model. 
On the string theory side, one obtains a very similar particle model
by an appropriate light cone gauge.
After obtaining and diagonalising the S-matrix of the particles
one can write down the Bethe equations for periodic states.

The particle model consists of $8$ bosons and $8$ fermions
above the half-BPS vacuum \cite{Berenstein:2002jq}.
These particles can be grouped in a $4\times 4$ matrix.
Then the rows transforms under a 
\[
\alg{h}=\alg{psu}(2|2)\ltimes \Reals^3
\]
superalgebra and the columns under a separate
$\alg{h}$ algebra 
with shared central charges \cite{Beisert:2005tm}.
The S-matrix therefore is a product of the S-matrices
for rows and for columns
\[
\smat^{\alg{psu}(2,2|4)}_{12}=S^{0,\alg{psu}(2,2|4)}_{12}\,
\smat^{\mathrm{bare}}_{12}\,\dot\smat^{\mathrm{bare}}_{12}.
\]
Consequently, it suffices to restrict to only one row of particles
and to find its S-matrix $\smat^{\mathrm{bare}}_{12}$. 
Remarkably, it turns out that the flavour structure of 
this S-matrix is completely determined by 
its $\alg{h}$ symmetry \cite{Beisert:2005tm}.
Moreover, the Yang-Baxter relation is automatically satisfied
by this S-matrix.
Symmetry alone, however, does not constrain the overall phase factor
$S^{0,\alg{psu}(2,2|4)}_{12}$.

The properties of the particles 
were mainly derived from gauge theory,
but also in string theory the particles behave similarly 
\cite{Hofman:2006xt,Arutyunov:2006gs}. 
It is therefore very reasonable to assume 
that the flavour structure of the S-matrix 
is the same for both models. 
Finding the exact phase factor for $\superN=4$ gauge theory
and for string theory on $AdS_5\times S^5$ remains one of the
biggest challenges in this context.
If the Bethe equations with the correct phase turn out to 
apply even at finite coupling, one could compare them
and see whether the AdS/CFT prediction
of coinciding spectra holds.

\section{Phase Factor and Crossing}

To obtain the phase factor for an integrable model
one usually employs crossing symmetry
which puts severe restrictions on its form.
Furthermore, assuming a minimal set of singularities
often fixes the factor uniquely.
An equation for crossing symmetry of the phase factor 
for the $\alg{h}$-symmetric S-matrix
$\smat^{\alg{h}}_{12}=S^{0,\alg{h}}_{12}\,\smat^{\mathrm{bare}}_{12}$
was derived by Janik in \cite{Janik:2006dc}
\[\label{eq:crossh}
S^{0,\alg{h}}(1/\xpm{1},\xpm{2})=
\frac{f(\xpm{1},\xpm{2})}{S^{0,\alg{h}}(\xpm{1},\xpm{2})}
\]
with the function
\[\label{eq:crossingfn}
f(\xpm{1},\xpm{2})
=\frac{\xm{1}-\xm{2}}{\xm{1}-\xp{2}}\,
\frac{\xp{1}-1/\xm{2}}{\xp{1}-1/\xp{2}}
=\frac{\xp{1}-\xp{2}}{\xm{1}-\xp{2}}\,
\frac{1/\xp{1}-\xm{2}}{1/\xm{1}-\xm{2}}\,.
\]
The spectral parameters $\xpm{}$ are related to the particle
momenta by
\[\label{eq:momdef}
\exp ip=\frac{\xp{}}{\xm{}}\,.
\]
They furthermore obey the equation
\[\label{eq:xpmnorm}
\xp{}+\frac{1}{\xp{}}-\xm{}-\frac{1}{\xm{}}=\frac{i}{g}\,,
\]
where $g$ is the square root of the 't~Hooft coupling constant 
(up to factors)%
\footnote{To simplify many expressions, I have chosen  
a normalisation in this letter which differs from 
my previous papers, e.g.~\cite{Beisert:2004ry}.
The relationship to the old literature 
is $\xpm{}=\xpm{\mathrm{old}}/\sqrt{2}\,g\indup{old}$ and
$g=g\indup{old}/\sqrt{2}$.
My excuses for all the previous factors of $1/\sqrt{2}$\ldots}
\[\label{eq:gnorm}
g=\frac{\sqrt{\lambda}}{4\pi}\,.
\]
The function $f$ has singularities at 
\[\label{eq:zerospoles}
\xpm{1}=\xpm{2},1/\xpm{2}\quad\mbox{and when}\quad
\xp{1}=1/\xm{2}\quad\mbox{or}\quad
\xm{1}=\xp{2}.
\]
The latter two singularities are 
related to two-particle bound states, c.f.~\cite{Dorey:2006dq}.

In fact, the crossing relation \eqref{eq:crossh} superficially has no 
solution due to the mismatch of both sides under the antipodal map 
$\xpm{1}\mapsto1/\xpm{1}$.
However, we have to take into account that the phase
factor $S^{0,\alg{h}}$ has branch cuts, 
e.g.~from the opening-up of a two-particle channel.
We therefore need to specify how to reach 
the antipodal point $1/\xpm{1}$ from the 
point $\xpm{1}$ itself:
The equation for the spectral parameters \eqref{eq:xpmnorm} 
defines a complex torus \cite{Janik:2006dc}
and there are at least two inequivalent short paths to reach the antipode.
Depending on which cuts are crossed by the path, 
the phase factor 
$S^{0,\alg{h}}(1/\xpm{1},\xpm{2})$ 
can take different values.
One of them should obey \eqref{eq:crossh},
the other will obey a related equation.

Going back to $\alg{psu}(2,2|4)$
we can express the crossing relation
in terms of the dressing factor $\sigma$ \cite{Arutyunov:2004vx} 
in the conventions of \cite{Beisert:2005fw} 
($\sigma$ in \cite{Arutyunov:2004vx} equals $\sigma^2$ in \cite{Beisert:2005fw})
as
\[
\label{eq:crossing}
\sigma(1/\xpm{1},\xpm{2})=\frac{h(\xpm{1},\xpm{2})}{\sigma(\xpm{1},\xpm{2})}\,,
\quad\mbox{where}\quad
h(\xpm{1},\xpm{2})=\frac{\xm{2}}{\xp{2}}\,\frac{1}{f(\xpm{1},\xpm{2})}\,.
\]
In this letter we shall propose a function 
obeying the above crossing relation.
This also leads to a function obeying crossing \eqref{eq:crossh} 
for the S-matrix with $\alg{h}$ symmetry via
\[
S^{0,\alg{h}}(\xpm{1},\xpm{2})=
\sqrt{\frac{\xp{1}-\xm{2}}{\xm{1}-\xp{2}}\,
      \frac{1-1/\xm{1}\xp{2}}{1-1/\xp{1}\xm{2}}}\,
\frac{1}{\sigma(\xpm{1},\xpm{2})}\,.
\]
%

\section{The Proposal}

The central proposal of this letter 
is that a crossing-symmetric phase factor is given by 
\<
\label{eq:phase}
\sigma(\xpm{1},\xpm{2})\eq\exp i\theta(\xpm{1},\xpm{2}),\nln
\theta(\xpm{1},\xpm{2})\eq
\theta_0(\xpm{1},\xpm{2})+\theta_1(\xpm{1},\xpm{2}),
\nln
\theta_0(\xpm{1},\xpm{2})\eq
-\frac{i}{2}\log\lrbrk{\sqrt{\frac{\xp{1}\xm{2}}{\xm{1}\xp{2}}}\,
\frac{1-1/\xm{1}\xp{2}}{1-1/\xp{1}\xm{2}}},
\nln
\theta_1(\xpm{1},\xpm{2})\eq
\sum_{r=2}^\infty
\sum_{s=r+1}^\infty
c_{r,s}\bigbrk{q_r(\xpm{1})\,q_s(\xpm{2})-q_s(\xpm{1})\,q_r(\xpm{2})}
\>
with the coefficients 
\[\label{eq:coeffs}
c_{r,s}=\frac{(-1)^{r+s}-1}{\pi}\,\frac{(r-1)(s-1)}{(r+s-2)(s-r)}\,
\]
and the magnon charges
\[\label{eq:magcharge}
q_r(\xpm{})=\frac{i}{r-1}\,\lrbrk{\frac{1}{(\xp{})^{r-1}}-\frac{1}{(\xm{})^{r-1}}}.
\]
The first contribution $\theta_0$ 
is similar at leading order in strong coupling 
with classical string theory \cite{Kazakov:2004qf} 
and the AFS phase \cite{Arutyunov:2004vx},
but it does not literally agree.
It is shown in a subsequent publication \cite{Beisert:2006ib}
that an additional homogeneous solution of the crossing relation \eqref{eq:crossh}
with $f=1$ is required
\[\label{eq:lcpiece}
\delta \theta(\xpm{1},\xpm{2})=
\frac{i}{2}\lrbrk{\frac{1}{2}+\frac{ig}{\xp{2}}-\frac{ig}{\xm{2}}}\log\frac{\xp{1}}{\xm{1}}
-\frac{i}{2}\lrbrk{\frac{1}{2}+\frac{ig}{\xp{1}}-\frac{ig}{\xm{1}}}\log\frac{\xp{2}}{\xm{2}}\,.
\]
Together, the two contributions $\theta_0+\delta \theta$ 
yield precisely the phase factor derived from 
string theory in light cone gauge \cite{Frolov:2006cc}
which is known to be consistent with the classical result
\cite{Kazakov:2004qf,Arutyunov:2004vx}.
The second contribution $\theta_1$ is precisely the
one-loop phase proposed by Hern\'andez and L\'opez \cite{Hernandez:2006tk}.
It was already confirmed that this phase obeys the
crossing relations perturbatively up to the first 
order in the strong-coupling expansion in \cite{Arutyunov:2006iu}.
The present claim is that the phase \eqref{eq:phase}
obeys the crossing relation exactly even for finite coupling $g$. 
In the following I will provide arguments to substantiate the proposal.
I will however not give a rigorous proof.

\section{Analytic Structure}

To motivate why the phase \eqref{eq:phase,eq:coeffs,eq:magcharge}
might receive no further corrections to be able to solve the crossing relation,
it is useful to investigate the structure of zeros and poles in
the function $\exp i\theta$.
The key insight is that, effectively, 
an almost analytic function is uniquely defined (up to a constant term) 
by the positions and multiplicities of its zeros and poles.
We would therefore need to show that the structure of poles
agrees with the crossing relation.
To find zeros and poles, we sum up the expressions
for $\theta_1$ \eqref{eq:phase,eq:coeffs,eq:magcharge} in a closed form,
see \cite{Arutyunov:2006iu}.
The phase then has the following general form
\[\label{eq:thetasummed}
\theta_1(\xpm{1},\xpm{2})=\ldots
\pm\frac{1}{2\pi}\log(\ast)\log(\ast)+\ldots
\pm\frac{1}{2\pi}\Li_2(\ast)+\ldots
\]
with $\ast$ representing some combinations of $\xpm{1,2}$.

Let us first consider the crossing relation applied twice:
It has the form 
\[\label{eq:doublecrossing}
\sigma(\xpm{1},\xpm{2})=\frac{f(1/\xpm{1},\xpm{2})}{f(\xpm{1},\xpm{2})}\,
\sigma(\xpm{1},\xpm{2})
\qquad\mbox{or}\qquad
\exp i\theta=\exp i\oldDelta\theta\,\exp i\theta.
\]
It is understood that the two instances of $\sigma$ differ 
by a continuous path once around the imaginary period of the torus.
In other words, we are comparing the value of the dressing factor $\sigma$ on two
different Riemann sheets. In detail, the phase shift $\oldDelta\theta$ reads 
\[\label{eq:doublecrossingfn}
\exp i\oldDelta\theta=
\frac{\xp{1}-\xp{2}}{\xm{1}-\xp{2}}\,
\frac{\xm{1}-\xm{2}}{\xp{1}-\xm{2}}\,
\frac{1-1/\xm{1}\xp{2}}{1-1/\xp{1}\xp{2}}\,
\frac{1-1/\xp{1}\xm{2}}{1-1/\xm{1}\xm{2}}\,.
\]
Now it is important to know how the functions $\log$ and $\Li_2$ change when 
crossing a cut 
\[\label{eq:sheetchange}
\log z\mapsto \log z\pm 2\pi i,\qquad
\Li_2 z\mapsto \Li_2 z\pm 2\pi i\log z.
\]
Therefore, the expression for the phase shift is of the form 
\[
\oldDelta \theta_1=\ldots\pm i\log(\ast)+\ldots\pm 2\pi+\ldots
\]
describing zeros and poles in $\exp i\oldDelta\theta$. 
Excitingly, the coefficients in front of the 
$\log$'s are precisely $\pm i$ leading to a \emph{single} zero or pole,
in agreement with \eqref{eq:doublecrossingfn}.

The appearance of integer factors in an exponent gives the 
hint that a non-re\-nor\-ma\-li\-sa\-tion theorem may apply:
Further corrections of $\theta$ would be of the order $\order{1/g}$ or higher.
As the coupling constant is arbitrary, it would be hard
to obtain \emph{integer} coefficients in front of 
logarithmic singularities. 
We would most likely introduce
zeros or poles with irrational weights which would completely
alter the analytic structure and introduce unwanted branch cuts.
This is similar to the argument of why some anomalies 
in a quantum field theory receive corrections at the one-loop level only.
In fact, as $\theta_1$ is a one-loop contribution, 
it is conceivable that it represents an anomaly of some sort. 
A further supporting argument in the form of an explicit example is given in \appref{sec:stack}.

Next we should find out where the logarithmic singularities
in $\oldDelta\theta_1$ reside. 
According to \eqref{eq:sheetchange},
the positions are just the zeros and poles of the arguments 
of $\log$'s and $\Li_2$'s in \eqref{eq:thetasummed}.
For the full expression from \eqref{eq:phase,eq:coeffs,eq:magcharge}, 
the logarithmic singularities happen to be at the positions
\[\label{eq:polepos}
\xexp{}{1}=\xexp{}{2},1/\xexp{}{2},\quad
\quad\mbox{or}\quad
\xexp{}{1,2}=0,\infty,\pm 1,
\]
where $\xexp{}{1,2}$ represent any of $\xpm{1,2}$.
Among these we clearly identify all the poles and zeros of the
function \eqref{eq:doublecrossingfn}.
Therefore it is conceivable that \eqref{eq:phase,eq:coeffs,eq:magcharge} 
solve the doubled crossing relation \eqref{eq:doublecrossing}.

It turns out that $\theta_1$ alone 
solves the doubled crossing relation.
The additional term $\theta_0$ is required 
to obey the single crossing relation \eqref{eq:crossing}.
Using the perturbative machinery of \cite{Arutyunov:2006iu}
it is straight-forward to verify that the crossing relation 
is obeyed for several odd powers in $1/g\sim\zeta$. 
In fact, we used this property to construct and guess the
phase $\theta_0$.
Conversely, the even powers in $1/g$ are 
generated only by $\theta_1$. One might be able to 
show that crossing is obeyed perturbatively for even powers
in $1/g$, but this would be much more involved due to
residual logarithms in the expressions.

Should \eqref{eq:phase,eq:coeffs,eq:magcharge} be the 
correct physical answer for string theory?
Firstly, as shown above, it has about the right set of analytic properties. 
Furthermore, it has been verified to obey the crossing relation in the leading 
and sub-leading perturbative orders \cite{Arutyunov:2006iu} 
(by construction, also to second and a few higher even orders).
Finally, further corrections of $\theta$ would be 
likely change the analytic structure and 
a non-renormalisation theorem may apply.
In conclusion, one of the simplest conceivable solutions of the crossing 
relation may be \eqref{eq:phase,eq:coeffs,eq:magcharge}.
Together with the homogeneous solution \eqref{eq:lcpiece} \cite{Beisert:2006ib}
it agrees with perturbative string computations \cite{Callan:2004uv,Frolov:2003xy}
at next to leading order \cite{Arutyunov:2004vx,Beisert:2005bv,Schafer-Nameki:2005tn,Hernandez:2006tk,Freyhult:2006vr}.
However, it must be noted that the proposed phase 
does not match with the classical scattering phase for 
giant magnons derived in \cite{Hofman:2006xt}.
A further homogeneous solution of the crossing relation 
may be required for agreement with giant magnons and/or 
higher perturbative orders.

\section{Verifications}

Of course we need to perform some basic tests of the conjecture.
The perturbative test in \cite{Arutyunov:2006iu} is a first step. 
We would like to show at finite coupling $g$ 
that the crossing relation \eqref{eq:crossing} is obeyed.
The problem is that we need to specify a path 
when shifting one of the parameters $\xpm{}$ 
to its antipode $1/\xpm{}$ and that 
the change of phase does depend on it.
A full analysis is beyond the scope of the current publication 
and should be performed in more detail elsewhere.

Here we are moderate and check the validity of the crossing relation
for a few random values of $\xpm{1,2}$. 
We specify a path $\xpm{1}(t)$ connecting
$\xpm{1}$ to $1/\xpm{1}$ and obeying \eqref{eq:xpmnorm}.
We use a parametrisation where the rapidity $z$ moves along one 
imaginary period of the torus
which defines the space of solutions $\xpm{}$ to \eqref{eq:xpmnorm},
c.f.~\cite{Janik:2006dc}. 
In particular, we choose the momentum $p$, defined via \eqref{eq:momdef}, 
to equal $p=2\ellAM z$.
Here `$\ellAM$' represents Jacobi's elliptic amplitude 
with elliptic modulus $k=4ig$. 
The imaginary half-period is
$\omega_2=2i\ellK(\sqrt{1-k^2})-2\ellK(k)$.
When moving along the path, we need to be careful 
about branch cuts. When the path crosses a branch cut,
the logarithms have to be replaced according to 
\eqref{eq:sheetchange}.

Several sets of points $\xpm{1,2}$ were chosen at random and the 
crossing relation \eqref{eq:crossing} 
was verified numerically to six-digit precision.
Furthermore, the double crossing relation
\eqref{eq:doublecrossing} can be verified more explicitly:
One simply keeps track of which terms 
have been added according to \eqref{eq:sheetchange}
when moving along the path. The resulting expression
matched \eqref{eq:doublecrossingfn}
(or its inverse, cf.~\cite{Beisert:2006ib})
in all tested cases.

\section{Conclusions and Outlook}

In this letter I have proposed an overall phase factor 
\eqref{eq:phase,eq:coeffs,eq:magcharge}
for the worldsheet S-matrix 
of strings on $AdS_5\times S^5$:
Its main property is that it seemingly solves Janik{}'s crossing relation
\cite{Janik:2006dc} stated in \eqref{eq:crossing} as the present tests confirm.
The S-matrix agrees with string theory at the classical level \cite{Kazakov:2004qf}
and the first subleading order in $1/\sqrt{\lambda}$ \cite{Hernandez:2006tk}
when the homogeneous piece \eqref{eq:lcpiece} \cite{Beisert:2006ib} is added.
It is useful to write down the overall scattering factor for two
excitations in the $\Reals\times S^3$ subsector
%
\[
\sigma^2(\xpm{1},\xpm{2})\,
\frac{\xp{1}-\xm{2}}{\xm{1}-\xp{2}}\,
\frac{1-1/\xp{1}\xm{2}}{1-1/\xm{1}\xp{2}}
=
\exp \bigbrk{2i\theta_1(\xpm{1},\xpm{2})}\,
\sqrt{\frac{\xp{1}\xm{2}}{\xm{1}\xp{2}}}\,
\frac{\xp{1}-\xm{2}}{\xm{1}-\xp{2}}\,
.
\]
Here $\theta_1$ is the Hern\'andez-L\'opez scattering phase
\cite{Hernandez:2006tk} reproduced in \eqref{eq:phase}.
It is interesting that such a simple form of scattering factor 
remains.

For the complete S-matrix element $A_{12}$ in the S-matrix
with $\alg{h}$ symmetry we obtain
\[
A_{12}=
S^{0,\alg{h}}(\xpm{1},\xpm{2})\,
\frac{\xp{2}-\xm{1}}{\xm{2}-\xp{1}}
=
\exp \bigbrk{-i\theta_1(\xpm{1},\xpm{2})}\,
\sqrt[\scriptstyle 4]{\frac{\xm{1}\xp{2}}{\xp{1}\xm{2}}}\,
\sqrt{\frac{\xp{2}-\xm{1}}{\xm{2}-\xp{1}}}\,
.
\]
Intriguingly, the element $D_{12}$
has just the square root term inverted
\[
D_{12}=-
S^{0,\alg{h}}(\xpm{1},\xpm{2})
=
-\exp \bigbrk{-i\theta_1(\xpm{1},\xpm{2})}\,
\sqrt[\scriptstyle 4]{\frac{\xm{1}\xp{2}}{\xp{1}\xm{2}}}\,
\sqrt{\frac{\xm{2}-\xp{1}}{\xp{2}-\xm{1}}}\,
.
\]
The term $A_{12}$ is precisely the square root of (the inverse of) the above
scattering factor for string theory in the $\Reals\times S^3$ sector. 
The full S-matrix for string theory can thus be written merely as
the product of two $\alg{h}$-symmetric S-matrices without an additional prefactor
(the inverse is due to a change of conventions)
\[
\smat^{\alg{psu}(2,2|4)}_{12}=\smat^{\alg{h}}_{12}\,\dot\smat^{\alg{h}}_{12}.
\]

There is a host of further investigations that should be performed:
The present conjecture should be completed by a 
suitable homogeneous solution of the crossing equation
to achieve full agreement with perturbative string theory at 
higher orders. 
Then it should be derived from string theory
along the lines of \cite{Arutyunov:2005hd,Frolov:2006cc,Klose:2006dd,Roiban:2006yc}. 
It is interesting to see that the one-loop result consists of 
$\log\cdot\log$ and $\Li_2$ terms. This is what might be 
expected as the outcome of a one-loop integral in some field theory
(albeit a four-dimensional one).

Furthermore, it is very important to study the analytic structure of the phase factor. 
Where are the zeros, singularities, branch points and how
are they connected? What is their meaning? 
Is there periodicity along the real cycle of the torus?
What is the structure of the underlying Riemann surface?
Does the function $\theta_1$ of two points on a torus
appear in another context?
Some of this knowledge should eventually enable one
to rigorously prove the crossing relation for the phase factor.

The most pressing question is presumably 
whether the phase factor interpolates to $\sigma=1$ 
at the first few orders of the weak-coupling expansion around $g=0$
in order to match with gauge theory. 
This is a crucial test of the AdS/CFT conjecture.
Simple agreement with gauge theory would very 
much count in favour of the exact validity of the correspondence.
In the case of disagreement, one may argue that the Bethe ans\"atze
are asymptotic and valid only to the first few orders in 
perturbation theory (at either strong or weak coupling)
\cite{Serban:2004jf}. 
Therefore the disagreement would be irrelevant
and both models would have their own dressing factor $\sigma$.
The author's hope, however, is that the Bethe equations 
for gauge and string theory are exact and not just asymptotic.
No tests of the weak-coupling regime have been performed here. 
It is only remarked that, according to the conventional logic,
the Hern\'andez-L\'opez term appears at $\order{g^3}$.
Perhaps, the correct gauge theory answer ($\sigma=1$ until at
least $\order{g^4}$) can be found on a different Riemann sheet?
In any case, the appearance of a contribution to 
anomalous dimensions at two \emph{and a half} loops 
(equivalent to $\order{g^3}$ in~$\sigma$) would seem cumbersome. 
Results in the BMN limit \cite{Berenstein:2002jq,Beisert:2004hm},
for transcendentality counting \cite{Kotikov:2004er,Eden:2006rx}
and the similarity to the Hubbard spin chain \cite{Lieb:1968aa,Rej:2005qt}
suggest that the exact function $\sigma=1$ is preferable for several reasons.
This function however cannot extrapolate to the 
factor obeying crossing symmetry (unless there are
essential singularities). 

An important test for the consistency of some
Bethe equations is that they reproduce the correct number of
states of the underlying model. This is not an easy task but it should
be performed for the present model and some tests seems possible. 
Does the number of states change between strong and weak coupling?
A related question was posed in \cite{Minahan:2005jq}:
How does the $\Reals\times S^3$ sector of string theory 
transform to the (smaller) $\alg{su}(2)$ sector in gauge theory?

Finally, it would be remarkable if 
one could apply sigma model Bethe equations 
with an additional level of nesting to derive the complete phase.
This has already been demonstrated to work in several cases 
\cite{Mann:2005ab,Gromov:2006dh,Gromov:2006cq}
at the leading order in strong coupling
(similarly, for gauge theory in \cite{Rej:2005qt}).
Can we also derive the Hern\'andez-L\'opez phase in this fashion,
perhaps even in more general sigma models?

\paragraph{Note added.}

A previous version of this manuscript 
posted to \texttt{\href{http://arxiv.org/abs/hep-th/0606214v1}{arxiv.org}}
claimed full agreement of the phase 
\eqref{eq:phase,eq:coeffs,eq:magcharge}
with perturbative string theory at the leading few orders.
It was brought to my attention by J.~Maldacena that this is
not so for the case of classical giant magnons \cite{Hofman:2006xt}
and AFS \cite{Arutyunov:2004vx} at finite momentum.
R.~Hern\'andez and E.~L\'opez noted that it does not
even agree with AFS at small momentum,
i.e.~with classical spinning strings and near-plane wave strings. 
The correct statement is that it agrees for small momenta 
after adding the homogeneous piece \eqref{eq:lcpiece}
as found in a subsequent article \cite{Beisert:2006ib}.
To achieve agreement with classical giant magnons is more subtle,
cf.~also \cite{Beisert:2006ib}.
I am very grateful for their kind remarks
and also to S.~Frolov and A.~Tseytlin for similar statements.

\paragraph{Acknowledgements.}

I thank N.~Dorey, S.~Frolov, R.~Hern\'andez,
C.~Kristjansen, E.~L\'opez,
J.~Maldacena, T.~McLoughlin, M.~Staudacher 
and A.~Tseytlin
for discussions. 
This work is supported in part by the U.S.~National Science
Foundation Grant No.~PHY02-43680. 
Any opinions, findings and conclusions or recommendations expressed in this
material are those of the author and do not necessarily reflect the
views of the National Science Foundation.

\appendix
\section{A Spiral Staircase of Poles}
\label{sec:stack}

The double crossing relation \eqref{eq:doublecrossing}
requires that the degree of a pole in $\sigma$ changes
by one when going once around one period of the torus.
To gather some experience with functions of this type, let
us consider the simple example
\[
f(y)\sim\prod_{n=1}^\infty\lrbrk{\frac{y-n}{y+n}}^n.
\]
This function was constructed to have a zero of degree $n$ at $y=n$ 
for all $n\in\Integers$. It needs regularisation and 
we can easily evaluate the product by considering a multiple logarithmic 
derivative and then integrating back.
The result is of the form 
\[
f(y)\sim \exp S(e^{2\pi i y})
\qquad\mbox{with}\qquad
S(w)
=\frac{\Li_2 w+\log w\log(1-w)}{2\pi i}
\,.
\]
This function is regular at $w=1$ and has a logarithmic singularity at $w=0$.
When moving once around the point $w=0$ (and thus past the branch cut)
a zero or a pole appears at $w=1$. Performing another loop in the 
same direction will increase the degree of the zero or pole by one.
This property led to the conjecture that the Hern\'andez-L\'opez term
\cite{Hernandez:2006tk} might be sufficient to satisfy large parts
of the crossing relation: In the summed form by Arutyunov and Frolov 
\cite{Arutyunov:2006iu} it consists of only $\Li_2$ and $\log\cdot\log$ terms
with precisely the right coefficients.
Moreover, one can show that $\theta_1$ is
a sum of $\pm S(w)$'s with the $w$'s some suitable 
functions of $\xpm{1,2}$.

\bibliographystyle{nbshort}
\bibliography{n4phase}

\end{document}